\documentclass{article}

\usepackage{amsmath}
\usepackage{PRIMEarxiv}

\usepackage[utf8]{inputenc} 
\usepackage{hyperref}       
\usepackage{url}            
\usepackage{booktabs}       
\usepackage{amsfonts}       
\usepackage{nicefrac}       
\usepackage{microtype}      
\usepackage{cite}
\usepackage{amsmath,amssymb,amsfonts}
\usepackage{graphicx}
\usepackage{parskip}
\usepackage{relsize}
\usepackage{booktabs}
\usepackage{multirow}
\usepackage{caption}
\usepackage{array}
\usepackage{bbding}
\usepackage{pifont}
\usepackage{wasysym}
\usepackage{amssymb}
\usepackage{multirow} 
\usepackage{textcomp}
\usepackage{stfloats}
\usepackage{url}
\usepackage{verbatim}
\usepackage{cite}
\usepackage{lineno}
\usepackage[utf8]{inputenc}
\usepackage{CJKutf8}
\usepackage[linesnumbered,ruled]{algorithm2e}
\usepackage{algpseudocode}
\usepackage{hyperref}
\usepackage{graphicx}
\DeclareUnicodeCharacter{2212}{\ensuremath{-}}

\usepackage{alphabeta}
\usepackage[utf8]{inputenc}

\usepackage{microtype}

\title{Base Station Deployment under EMF constrain by Deep learning
\thanks{\textit{{
This work has been accepted to the IEEE for publication. Copyright may be transferred without notice, after which this version may no longer be accessible.}}: 
}}

\author{
{Mohammed Mallik\textsuperscript{1}, and Guillaume Villemaud\textsuperscript{1}} 
\\\\
\small
\textsuperscript{1}INSA Lyon, Inria, CITI, UR3720, 69621 (e-mail: firstname.name@insa-lyon.fr)\\
\\
}
\begin{document}
\maketitle

\begin{abstract}
As 5G networks rapidly expand and 6G technologies emerge, characterized by dense deployments, millimeter-wave communications, and dynamic beamforming, the need for scalable simulation tools becomes increasingly critical. These tools must support efficient evaluation of key performance metrics such as coverage and radio-frequency electromagnetic field (RF-EMF) exposure, inform network design decisions, and ensure compliance with safety regulations. Moreover, base station (BS) placement is a crucial task in the network design, where satisfying coverage requirements is essential. To address these, based on our previous work, we first propose a  conditional generative adversarial network (cGAN) that predicts location specific received signal strength (RSS), and EMF exposure simultaneously from the network topology, as images. As a network designing application, we propose a Deep Q Network (DQN) framework, using the trained cGAN, for optimal base station (BS) deployment in the network. Compared to conventional ray tracing simulations, the proposed cGAN reduces inference and deployment time from several hours to seconds. 
Unlike a standalone cGAN, which provides static performance maps, the proposed GAN-DQN framework enables sequential decision making under coverage and exposure constraints, learning effective deployment strategies that directly solve the BS placement problem. Thus making it well suited for real time design and adaptation in dynamic scenarios in order to satisfy pre defined network specific heterogeneous performance goals. 
\end{abstract}



\maketitle

\section{Introduction}
\label{sec:introduction}
The transition from 5G to 6G introduces both opportunities and complexities in coverage optimization. High-frequency technologies such as millimeter wave (mmWave) enable high capacity, low latency communication. Yet are constrained by pronounced path loss and limited propagation, particularly in dense urban environments. Consequently, ultra dense base station deployment is required to ensure reliable connectivity while adhering to International Commission of Non-Ionizing Radiation Protection (ICNIRP) guidelines on radio frequency electromagnetic field (RF-EMF) exposure, which remain integral to safe network design. In parallel, cell free massive multiple input multiple output (CF-MIMO) characterized by a distributed network of dense access points with few antennas each \cite{interdonato2019downlink}, is envisioned as a foundational technology for 6G, offering scalable and resilient coverage.

Conventional approaches to network optimization, based in mathematical modeling and operations research, often exhibit limited efficiency and scalability in realistic communication environments \cite{SKOCAJ2022403,gong,article}. The highly dynamic and heterogeneous nature of large scale wireless systems introduces uncertainties and nonlinear interactions that hamper the accuracy and generalizability of such models. Consequently, traditional methods frequently yield suboptimal solutions, even when complemented with heuristic or meta-heuristic algorithms, whose direct deployment in dynamic scenarios remains challenging \cite{9992177}. Within this context, the base station (BS) placement problem emerges as a central task in network design, as BS locations fundamentally determine achievable coverage across 5G and beyond \cite{8316776,8658363,10225848}. Formulated as a combinatorial optimization problem and classified as NP-hard \cite{arora2003approximation}, base station (BS) placement requires selecting optimal deployment sites under constraints such as terrain limitations and geometric distribution models \cite{yoon2025study,LI2025111431}. In addition, BS placement must comply with safety limits defined by the World Health Organization (WHO) and the International Commission on Non Ionizing Radiation Protection (ICNIRP) through established EMF exposure guidelines. The exponential growth in computational complexity with increasing candidate sites further underscores the necessity of developing advanced optimization techniques capable of addressing BS placement effectively and within practical time constraints  \cite{LIAQ2025103983,LI2023119224}. 

BS placement and network performance evaluation typically rely on manual network planning informed by empirical field measurements or on computationally intensive ray-tracing simulations \cite{egea2019vehicular}. These approaches provide high accuracy predictions. Though their inherent labor intensity and limited scalability render them inadequate for the ultra dense network architectures anticipated in 6G. Their static nature prevent dynamic reconfiguration of network topology. For instance, the rapid deployment of mobile base stations or performance optimization under scenarios of highly variable user demand, such as mass gatherings, disaster recovery operations, or military missions. In this context, for network performance evaluation and optimization under EMF constraints, machine learning (ML) is utilized as a powerful tool to extract relevant features from the mobile network data. Deep learning (DL) has achieved great improvements and extensive applications on network optimization and performance metric evaluation. There are few studies to analyze effective network performance metric prediction model and optimization of BS deployment for the purpose of achieving better coverage performance with EMF and lower overhead.

\section{Related Works}
This section presents different strategies that have been used to address the network performance metric evaluation and optimization from the network topology. We can differentiate different approaches for network performance estimation based on propagation models, deep learning and network optimization techniques.
\subsection{Simulators}  

Evaluating network performance in an urban environment, whether in terms of coverage or EMF exposure, remains a challenging and computationally intensive task. Traditional deterministic tools, such as Veneris-Opal \cite{egea2019vehicular} and InoWave-Siradel (https://www.siradel.com/fr/telecom/
inowave-modelisation-frequences/), rely on RT techniques. The open-source RT simulator, Veneris-Opal \cite{egea2019vehicular}, have gained attention for their reliable performance and comparatively fast computation times. In these methods, the propagating EMF is modeled as an array of rays that reflect, diffract, and scatter through the environment, using the high-frequency approximation of Maxwell’s equations (optical ray theory). While RT provides detailed insights, it requires simplifying real world propagation phenomena and environmental parameters, which limits its adaptability to dynamic scenarios. 
Despite their high computational cost and sensitivity to the accuracy of the three-dimensional environmental model \cite{yun2015ray}, RT-based approaches, along with empirical or semi-empirical models such as close-in (CI), floating intercept (FI), or alpha beta gamma (ABG) models \cite{6831690, 6363950, 339880}, are often used for calculating coverage in metropolitan areas \cite{rizk1997two, wahl2005dominant, zugno2020toward}. For practical assessment of cellular network performance, considering all active networks and devices is often infeasible. Among the few accessible solutions,  Sionna \cite{hoydis2023sionnartdifferentiableray} is popular due to their reliable performance and relatively fast computation times, making them suitable choices for urban network evaluation studies.

\subsection{Deep Learning}

In the design of mobile communication systems, the assessment of network performance indicators; particularly coverage and EMF exposure represent a fundamental requirement, especially in alignment with safety recommendations issued by WHO and the ICNIRP. To accelerate and enhance the accuracy of such evaluations, recent studies have increasingly turned to deep learning models.

For instance, Xu et al. \cite{7161398} investigated the problem of power spectrum mapping in urban cognitive radio networks. In that work, a convolutional neural network (CNNs) model with a generative adversarial network (GAN) was developed to predict power spectrum (PS) maps. The method was evaluated under the assumption of uniformly distributed users and tested with bandwidths of 25 MHz and 75 MHz. The GAN architecture, inspired by the autoencoder framework, reconstructed complete PS maps from under-sampled input data, with synthetic training data generated via an inverse polynomial law model.
Similarly, Zhuo et al. \cite{8794603} introduced a self supervised GAN based framework to infer reference RF maps from incomplete observations. In their work, reference maps for training were constructed using a K nearest neighbor algorithm. In addition to these approaches, several recent contributions \cite{saito2019two,chouvardas2016method,11148193} have applied deep learning for coverage prediction. These studies demonstrated that CNNs can effectively approximate radio maps corresponding to specific transmitter and receiver configurations. However, a critical limitation arises: the trained models relies on the underlying city topology, requiring retraining for each new urban scenario. Several works have been done using matrix completion for radio map estimation \cite{tognola2021use,10715541}, avoiding the training phase with reference maps and where kernel methods are not used. For example, Wang et al. \cite{wang2021fast} used matrix completion for low-ranked matrices to construct radio maps in an indoor system. 

Recent studies have explored the use of deep learning for EMF exposure prediction in both indoor and outdoor environments. In  \cite{mallik2022eme,mallik2022towards,mallik2023eme}, 2 approaches were investigated: a UNet-based CNN \cite{ronneberger2021u} and a conditional GAN (cAGN) \cite{mirza2014conditional}. Their models were trained using reference EMF exposure maps generated from RT simulators \cite{amiot2013pylayers,egea2019vehicular}. However, in these works, network topology information, such as base station positions was not included as input during the training phase.
Several studies focus on the prediction of uplink (UL) and downlink (DL) exposure from mobile devices. For UL, several studies \cite{mazloum2021artificial,falkenberg2018machine} used artificial neural network (ANN) based models to estimate the emitted power of mobile phones, while others \cite{tognola2021use,10715541} focused on forecasting DL exposure. These approaches typically rely on readily available features, such as DL signal quality indicators (e.g., reference signal received power), combined with environmental descriptors.
However, the goal of the above mentioned work was only to predict EMF exposure. Thus, estimating RSS and EMF exposure simultaneously by a single deep learning model from its network topology was not considered.

\subsection{Optimization}
Numerous approaches have been used on base station (BS) location optimization. Heuristic and meta heuristic approaches were used in early works, frequently in conjunction with optimization frameworks and computational geometry \cite{9079581, 8788518, 10257851}.  For instance, \cite{10257851} presented an immune genetic algorithm that assessed performance based on strong and poor coverage regions. In \cite{8788518}, used a geometry driven genetic algorithm (GA) to solve BS placement. Similarly, \cite{9079581} suggested a two phase optimization approach. The first stage estimated RSS from the network topology using a regression model, and the second stage used the trained model to predict coverage, directing the deployment of heuristic-based BS (similar to the proposed work).  In addition to heuristics, a number of theoretical strategies have been proposed to address the BS deployment challenges' intrinsic non convexity and combinatorial nature \cite{7161398, 8119520, 8292392}.  
 In addition, DRL algorithms have been mainly used for aerial BS placement to operate alone or to support terrestrial network infrastructure to improve users coverage and throughput. For instance, a single deep Q network (DQN) 
agent is used to control aerial BSs \cite{8717805, aaa, ZHAO2025109844,electronics10232953,9149258}. A multi-agent RL (MARL) approach is utilised to control multiple aerial BSs \cite{ZHAO2025109844}. In these works, the Deep RL (DRL) has been mainly used for aerial BS location placement, while our work considers street-level base stations. In another interesting work \cite{samal}, authors have considered the use of DRL to maximize coverage. In that work, the optimizing parameter was antenna tilt while a groups of user clusters was distributed randomly.

Addressing the challenges of high frequency network optimization and motivated by the above mentioned works, this study proposes GAN-DQN, base station deployment framework that uses cGANs with deep Q-learning. Central to this framework is NPE-GAN (Network Performance Estimator Generative Adversarial Network), an enhancement of our previous Auto-RSS model \cite{auto-rss}. NPE-GAN uses a CNN-based conditional adversarial digital twin (DT) to predict location specific RSS and EMF exposure. These predictions allow the system to compute rewards that simultaneously capture coverage and EMF exposure in that scenario. Through reinforcement learning strategies, GAN-DQN significantly reduces computational complexity and time when compared to heavy heuristic methods. Proposed method offers a more scalable, EMF exposure aware, and efficient solution, as a practical candidate for real-time network optimization.

\subsection{Contributions}
We introduce GAN-DQN, a BS deployment strategy by utilizing DQN and a novel generative DT model, NPE-GAN. The key contributions are summarized below:
\begin{itemize}
    \item \textbf{NPE-GAN}: Based on our previous work AUTO-RSS \cite{auto-rss}, We present a generative conditional adversarial network (cGAN) - NPE-GAN that accurately predicts location specific network performance metric, such as RSS and EMF exposure.

    \item \textbf{GAN-DQN} presents a DRL framework for BS deployment that integrates NPE-GAN to provide rapid and accurate reward evaluations. By using NPE-GAN predictions, GAN-DQN can efficiently optimize network coverage with EMF exposure constraints to learn deployment strategies that balance performance and safety, and adjust dynamically to varying network conditions.
    \item Results shows that, NPE-GAN shows minimal error in prediction and by using NPE-GAN, the DQN model reduces simulation time from hours to seconds compared to Brute Force search and other search algorithm in BS deployment scenario, making it practical for real time optimization.
\end{itemize}

The paper is organized as follows. Section \ref{method} presents the proposed methodologies from NPE-GAN and GAN-DQN. Section \ref{training_setup} describes the training parameters and dataset generation process. Section \ref{evaluations} reports the experimental results, covering the dataset, baseline models, and performance evaluation. Finally, Section \ref{dis} and \ref{conclude} provides the discussion and conclusion.

\section{Methodology}\label{method}
In this Section in \ref{NPE-GAN}, the methodologies of EMF exposure estimation by NPE-GAN and DQN based optimization method GAN-DQN in \ref{gandqn} and \ref{DQN_GAN} are described.

\subsection{Network performance metric estimation by conditional GAN}\label{NPE-GAN}
This section describes the method to predict RSS and EMF exposure by a conditional GAN Model.

\subsection{NPE-GAN}
\begin{figure*}
    \centering
    \includegraphics[scale=0.55]{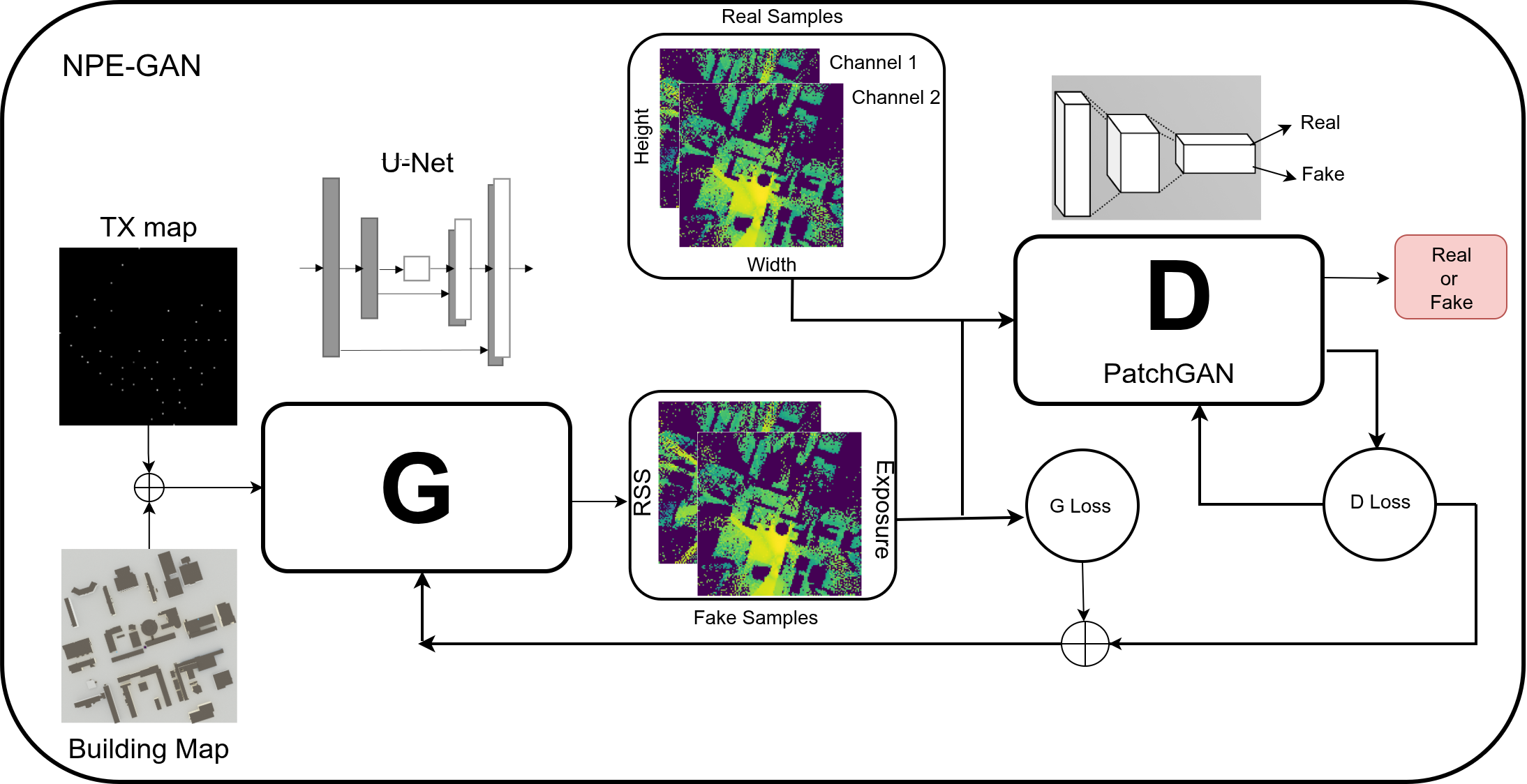}
    \caption{NPE-GAN: a cGAN model to predict network performance metrics. The input of the model is the binary map of TX and building topology and the output is the RSS and EMF exposure.}
    \label{modelgan}
\end{figure*}

In our previous work, by Auto-RSS \cite{auto-rss}, RSS was predicted in Lyon, France, from its topology using a convolutional autoencoder (CNN-AE). In this work we extend the concept further by developing a novel cGAN architecture capable of predicting location specific RSS and EMF exposure in a given environment. Similar to the environment model in \cite{auto-rss}, in this work, a $800 \times 800$$m^2$ area in INSA Lyon university is chosen. The innovation lies in the model architecture to handle the data structure and predict RSS and exposure simultaneously from the network topology. A high level overview of the proposed model is depicted in fig. \ref{modelgan} and detailed input and output data dimensions are provided in Sec. \ref{dataproc}.

\subsubsection{Network Architecture}\label{archi}
The proposed NPE-GAN consists of a UNet architecture for the generator and the PatchGAN architecture for the discriminator. Below, we provide a detailed description of both components and their architectures, elucidating the operational mechanisms that facilitate the accurate translation
from BS locations to the network performance metrics.

\begin{figure}[t]
    \centering
    \includegraphics[scale=0.60]{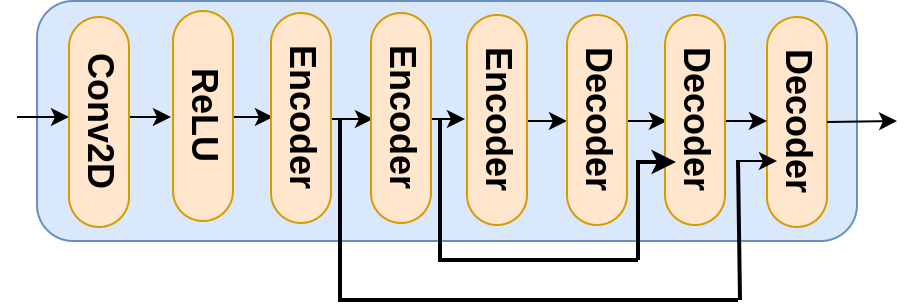}
    \caption{NPE-GAN: UNet Generator architecture}
    \label{unet}
\end{figure}
\subsubsection{Generator}
The generator uses a UNet architecture \cite{ronneberger2021u}, depicted in Fig. \ref{unet}, since this kind of network has proved to be very effective at maintaining spatial alignment when performing image to image translations. This is an essential property of a model developed for our problem, because both the input (building and BS location maps) and output (coverage and exposure maps) are given as rasterized grids. The input tensor has dimensions
$N \times N \times 2$, and the generator predicts a corresponding $N \times N \times 2$. 
Each encoder block applies a convolution (kernel size 4, stride 2), followed by ReLU activation and batch normalization. Decoder blocks mirror this structure using transposed convolutions (kernel size 4, stride 2), ReLU activations, and batch normalization, except for the final decoder, which uses a  $g(x) = tanh(x)$ activation to constrain outputs to the range 
$[-1, 1]$. To maintain spatial information, skip connections concatenate feature maps from encoder layer $i$ to decoder layer $n−i$, enabling the generator to combine high level semantics with fine spatial cues.



\begin{figure}[h]
    \centering
    \includegraphics[scale=0.60]{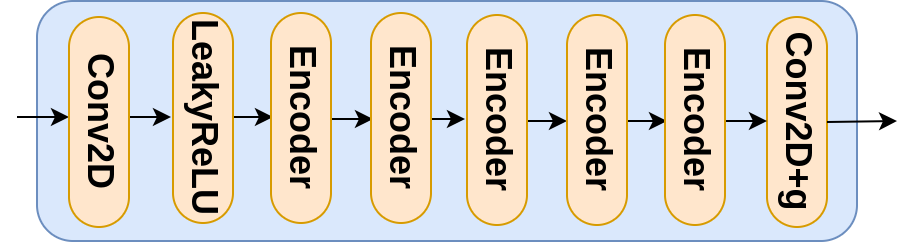}
    \caption{NPE-GAN: PatchGAN Discriminator architecture}
    \label{patch}
\end{figure}
\subsubsection{Discriminator: PatchGAN}
The discriminator follows the PatchGAN design \cite{demir2018patchbasedimageinpaintinggenerative}, illustrated in Fig. \ref{patch}. Instead of producing a single real or fake classification, PatchGAN classifies overlapping $70 \times70$ patches of the input image, which stabilizes training and improves local structure. The discriminator processes the concatenated input–output through a convolution block, followed by multiple encoder blocks with kernel size 4, and stride 2. Its output is a 2D activation map whose entries represent per patch realism probabilities. This architecture is resolution independent and can accommodate inputs larger or smaller than 
$128 \times128$ without modification. 




\subsection{Loss Function}

The architecture of cGAN is divided into two core components: the generator $(G_{\omega})$ and the discriminator $(D_{\tau})$, where $\omega$ and $\tau$ denote the neural network parameters associated with each. 

The loss function of the proposed cGAN model contains the discriminator and the generator as shown in (1):
\begin{equation}\label{eq:1}
\begin{aligned}
\mathcal{L}_{cGAN}(\omega,\tau) =\;
& \mathbb{E}_{x,y}[\log D_{\omega}(x,y)] \\
& + \mathbb{E}_{x,z}[1 - \log D_{\omega}(x,G_{\tau}(x,z))]
\end{aligned}
\end{equation}
where expectations are over the joint distribution of (x, y). This loss function essentially captures the discriminator’s probability assessment of y being the ground truth output
for a given x. 

The generator $G$ is not only trying to reduce the loss from the discriminator but also trying to move the fake distribution close to the real distribution by using $L1$ loss, which is given in 
\eqref{eq:2}:
\begin{equation}\label{eq:2}
    \mathcal{L}_{L1}(G)={E_{x,y,z}}[\left\| y-G(x,z) \right\|]
\end{equation}
The loss function of the generator network is stated in 
\eqref{eq:3}:
\begin{equation}\label{eq:3}
    G^\star =\arg\min_{G}\max_{D} L_{cGAN}(G,D)\lambda_{L1}(G)
\end{equation}
To optimize the performance of the cGAN, the objective function incorporates both the cGAN loss and an $L1$ regularization term. The objective term is given below:
\begin{equation}\label{eq:4}
    G^\star =\arg\min_{\omega}\max_{\tau} \mathcal{L}_{cGAN}(\omega,\tau)+ \lambda_{L1}(\omega)
\end{equation}

cGAN model and training detailed hyperparameters are provided in Section \ref{training_setup}, Table \ref{cgan}.

\subsection{BS Placement optimization by DRL}\label{gandqn}
This section focuses on the BS placement optimization problem into a DQN framework. We explore how implementing DQN can enhance communication services in a designated geographic region by using reinforcement learning techniques.
\subsection{Problem Formulation and DQN approach}\label{dqnprob}
In this work, the objective is to maximize coverage by strategically deploying base stations (BSs) to satisfy specific coverage requirements while considering EMF exposure limits within a given network. To compute coverage, the target region is uniformly rasterized into $L\times L$ pixels. For example, an area of size 
$800m \times 800m$ can be discretized into a 
$128 \times 128$ pixels. At each pixel location $(i,j)$, the RSS and EMF exposure are predicted by the NPE-GAN model, denoted as 
 $rss_{(i,j)}$ and $exp_{(i,j)}$, respectively. The goal is to identify BS placements that maximize the coverage rate $CR$. This is defined as the fraction of the area where the received power $rss_{(i,j)}$ exceeds a specified threshold  $\phi$, thereby ensuring acceptable Quality of Service (QoS). Let the BS locations be represented by the coordinates 
$(i, j)$ on the site-specific grid. The corresponding objective function to be maximized can then be expressed as:

\begin{equation}\label{eq5}
\begin{aligned}
&\underset{(i,j)}{\max} \sum_{(i,j) \in \mathcal{R}} CR, \\
&\text{s.t.} \quad C1:   \{i, j\} \in \mathcal{B}, \\
& \quad \quad \ C2:   {ER} \leq E_{th} 
\end{aligned}
\end{equation}
 where $\mathcal{R}$ denotes the region where RSS and exposure can be measured and users can be located, such as non building areas, $ER$ is the exposure rate which set to a threshold $E_{th}$ for feasibility, above that, the BS locations are taken as non deployable and the set $\mathcal{B}$ is the deployable locations available. 

\subsection{NPE-GAN for reward calculation}\label{gan_rew}

In deep reinforcement learning, reward design is crucial. Complex network simulations to evaluate the performance of a network is also time consuming. Traditional RT is needed to calculate network performance from base station placements, which can take minutes, making it much too slow for real time training. For quick reward evaluation, GAN-DQN uses NPE-GAN. Site specific coverage and EMF exposure maps are generated in milliseconds by the cGAN model.  This enables efficient evaluation of BS deployment scenarios by allowing the DQN agent to rapidly evaluate performance following each action. NPE-GAN speeds up training and enables GAN-DQN to take actions fast, offering scalable, nearly instantaneous feedback.  In the following section, the GAN-DQN approach is described.

\subsection{Placement optimization by DQN}\label{DQN_GAN}
We propose GAN-DQN, a framework that integrates a DQN RL agent to address the challenge of BS deployment in 5G and beyond. The DQN algorithm is particularly appropriate for this task because it effectively learns by exploration trying out new candidate deployment positions, and exploitation, refining strategies based on past successful placements. The optimization objective focuses on maximizing coverage while considering EMF exposure constraints. In addition, the scenario accounts for pre-existing BS deployments, reflecting a practical situation where infrastructure already partially exists. The agent then learns how to strategically introduce one or more additional BSs to enhance coverage without exceeding EMF exposure thresholds. 
\subsection{Deep Q Network}
In the GAN-DQN framework, the challenge of base station (BS) deployment is formulated as a markov decision process (MDP). This formulation allows an agent to explore and interact with the environment actively. Ultimately learning strategies that lead to optimal BS placement.

\subsubsection{MDP Design}
\subsection*{Environment}
In reinforcement learning (RL), an agent interacts with the environment and observes the rewards or penalties that ensue, improving its decisions iteratively.  The agent's goal is to optimize the location of base stations (BS) while accounting for EMF exposure levels and coverage area. The state $s_t$ from the state space $S$, which encodes the deployment specific network topology. This includes building layouts, terrain, and, if relevant, pre deployed BS sites represented as a binary image and coverage state are seen by the DQL agent at a specific time step $t$.  The agent uses its policy $\pi(a_t|s_t)$, which specifies the likelihood of selecting a specific action given the current state, to choose an action $a_t$ from the action space $A$ based on this state. Upon executing action $a_t$, the environment provides a reward $r_t$ and transitions to a new state $s_{t+1}$, reflecting the updated network configuration and performance after the placement of the new BS.

The agent restarts after continuing the process until it reaches the terminal state. The agent's objective is to optimize the discounted accumulated reward, which is described as
\begin{equation}
    R_t =\sum_{k=0}^{\infty} \gamma^{k}r_{t+k},
\end{equation} 
Here, $\gamma \in (0, 1]$ is the discount factor which determines the importance of future rewards compared to current reward. An action value function given in \eqref{rew}:
\begin{equation}\label{rew}
Q\pi(s, a) = E[R_t|st = s, at = a]     
\end{equation}
which is the expected return for selecting action $a$ in state $s$ and then follow a policy $\pi$. An optimal action-value function $Q \ast (s, a) = max_\pi Q_\pi(s, a)$ is the maximum action value achievable by following any policy for state $s$ and action $a$. The optimal action value function can be expressed by the Bellman equation as follows:
\begin{equation}
Q{\ast}(s, a) = E_{s^{\prime}} [r + \gamma \max_{a} Q \ast({s_0}^{\prime} , {a_0}^{\prime})|s, a].    
\end{equation}

\subsection*{Action}
The agent operates within an action space $A$, which comprises all candidate locations for BS deployment. At a given time step $t$, the agent selects an action
\begin{equation}
a_t = (i, j), \quad {i, j} \in \mathcal{A}
\end{equation}
where $(i,j)$ denotes the coordinates of the BS placement on the site-specific map, and $\mathcal{A}$ represents the set of all permissible deployment sites. BSs are placed sequentially, at each step, the agent chooses a new location informed by the current network state and previous placements. This sequential decision making enables efficient computation and allows the agent to adapt dynamically as it learns from earlier actions. We also used action masking such that only new available locations are chosen, despite the previously placed BSs or pre-deployed ones.

\subsection*{State}
The state $s_t$ at time step $t$ is fundamental for representing details of the network environment, serving as the key factor that shapes agents’ decision making within the optimization framework. It encodes communication related information for the grid inside the RoI. Formally, the state is
expressed as:
\begin{equation}
    s_t = {\mathcal{S}},
\end{equation}
where, $\mathcal{S}$ is the site specific grid information.  \\
 
\textit{Grid Representation}
The target region is discretized into an $m \times n$ grid. We represent the state using a tensor of size $2 \times m \times n \times n$.
Each of the features encodes distinct information for the BS deployment scenario: Coverage state of the grid/cell. Indoor/outdoor status of each cell, captured as a binary value (0 or 1), where 0 indicates an indoor location. This corresponds to the building map, all these from the preceding time step are considered, culminating in forming the tensor.

This structured representation allows simultaneous encoding of environmental features and network deployment in a single tensor as the state at each time step.

\subsection*{Reward}

As discussed in Sec.~\ref{gan_rew}, calculating the reward after every BS deployment is required for the evolving network conditions by evaluating both coverage and EMF exposure in the RoI. This is a computationally expensive task. The GAN−DQN addresses this bottleneck by the NPE-GAN model. NPE-GAN calculates rapid and precise estimation for coverage and EMF exposure. This approach ensures that the reward computation remains both accurate and computationally tractable, allowing the training process to proceed in a realistic yet efficient manner.

In this work, the objective is to maximize coverage rate with EMF exposure constraints. It is important that the reward $r_t$ at each time step $t$ reflects the optimization problem. To achieve this, we have chosen to maximize the $CR$. Therefore, our reward function $r_t$ is given as:
\begin{equation}
    r_t = 
    \begin{cases}
      CR & \text{if ER $\geq$  $\lambda$} \\
      -0.1 & \text{otherwise}
    \end{cases} 
\end{equation}
where, $CR$ is the coverage rate, $ER$ is the exposure rate and the term $\lambda$ corresponds to $ER$ threshold (which is chosen to be 0.90). Although the exposure rate does not appear explicitly in \eqref{eq5}, it is implicitly incorporated through the reward formulation, ensuring that EMF exposure considerations are accounted for in the evaluation process. If the exposure exceeds the predifined threshold the reward reflects the coverage improvement. Otherwise, a penalty is applied to guide the DQN away from unsafe deployments. We introduce two binary indicator for coverage and EMF exposure, $\mathbb{C}_{(i,j)}$ and $\mathbb{U}_{(i,j)}$ defined as

\begin{equation}
     \mathbb{C}_{(i,j)} =
    \begin{cases}
      1 & \text{if $\mathit{rss_{(i,j)}}$ $\geq$  $\phi$} \\
      0 & \text{if $\mathit{rss_{(i,j)}}$ $<$  $\phi$}
    \end{cases} 
\end{equation}
\begin{equation}
    \mathbb{U}_{(i,j)} =
    \begin{cases}
      1 & \text{if $\mathit{exp_{(i,j)}}$ $\leq$  $\gamma$} \\
      0 & \text{if $\mathit{exp_{(i,j)}}$ $>$  $\gamma$}
    \end{cases} 
\end{equation}
where $\phi$ and $\gamma$ are predefined RSS and exposure thresholds.

The rate can be calculated for the network performance metrics for a specific area with several BSs, which is defined as the ratio of the number of grids greater or less than a certain threshold to the total number of girds, i.e., $X = \sum_{(i,j)}^{L}X_{i,j}/L$. $X$ can be $\mathit{rss_{(i,j)}}$, $\mathit{exp_{(i,j)}}$ or any other metrics. In this case, $\phi$ and $\gamma$ are -110 dBm and 70 dB$\mu$V/m. The EMF exposure threshold is set at that value, as the French exposure limit is 6 V/m, we adopt a lower value to provide a conservative safety margin and ensure stable training.

\subsection*{Loss function}
The agent is trained following the DQN algorithm \cite{mnih2015human}, enabling it to interact iteratively with the environment. At each step, the agent chooses BS deployment actions and receives rewards based on the resulting improvements in network coverage, constrained by the EMF exposure threshold. Through these interactions, the agent refines its policy, denoted as $\pi_\theta(a_t|s_t)$. 
Standard reinforcement learning techniques, however, are limited to problems with relatively small state and action spaces. To overcome this, a deep neural network is employed to approximate the $Q$ function. The training procedure incorporates experience replay, which allows the agent to sample transitions offline, and a target network to stabilize learning. Initially, the target network is a copy of the $Q$ network. During training, the target network parameters are updated less frequently than those of the $Q$ network, providing a stable reference for the $Q$ updates.

\begin{figure}[t]
    \centering
    \includegraphics[scale=0.6]{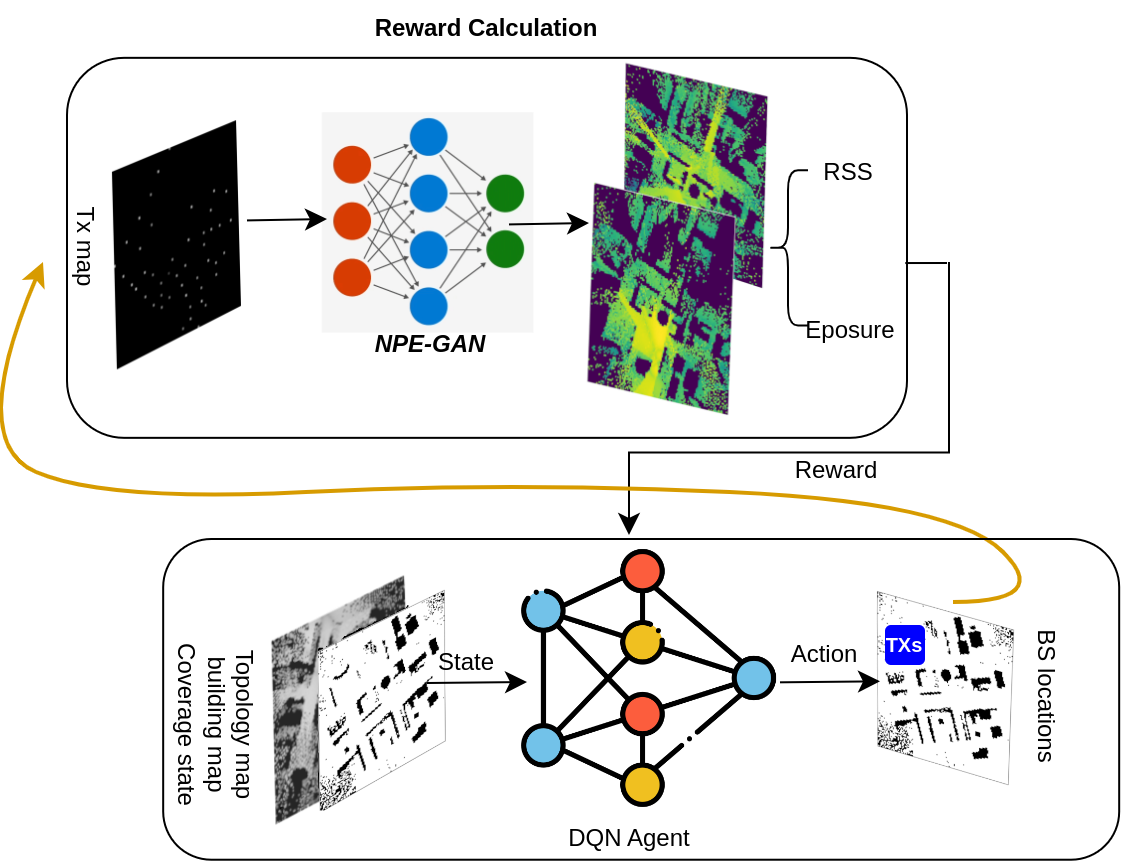}
     \caption{DQN training pipeline: with NPE-GAN providing  RSS and EMF exposure predictions to compute the reward at each
step}\label{traindqn}
    
\end{figure}

The DQN agent learns by minimizing the following loss function:\begin{equation}\label{loss}
    \mathcal{L}(\theta) = \mathbb{E}[(y_t - Q(s_t,a_t; \theta))^{2}],
\end{equation}
where $\theta$ are the weights of the DQN network, updated during training to approximate the optimal Q function. The policy $\eta$ is derived from this Q function, typically via an $\epsilon$-greedy strategy.  The target function $y_t$ is given as :
\begin{equation}
    y_t = r_t + \gamma \max_a Q(s_{t+1}, a_t, \theta_{target})
\end{equation}
A high-level overview of our DQN training is depicted in Fig. \ref{traindqn}.
\subsection*{Agent Model}\label{agentt}
The agent uses a deep Q network (DQN) architecture built upon a convolutional neural network (CNN). The CNN model receives as input an image like state representation of size
$128 \times 128 \times 2$.
The agent model architecture consists of the following components:
Input Layer: Receives the state tensor of dimension
$128 \times 128 \times 2$; Convolutional Layers:
The first convolutional block applies 16 filters of size $3\times3$ with ReLU activation and same padding. The second and third convolutional layers progressively have 32 and 64 filters, respectively, with ReLU activations and same padding. Max pooling with a $2\times2$ window downsamples the spatial dimensions. A fourth convolutional layer with 128 filters learns higher-level spatial features. 
The output layer is a fully connected layer with the same dimensionality as the size of the action space $\mathcal{A}$, using a linear activation in order to predict Q values for every possible action. For training, the model is compiled with the Adam optimizer and uses mean squared error (MSE) as the loss function, consistent with standard DQN architectures.
\subsection*{Proposed DQN training}
\begin{algorithm}
\SetKwInOut{Input}{Input}
\SetKwInOut{Output}{Output}

\Input{Tensor representation of the environment
obtained from the state with a binary building map, binary BS map}
\Output{Optimized BS locations}
Initialize replay memory D to a maximum capacity\;
Initialize action value function Q with random weights $\theta$\;
Initialize target network with weights $\theta_{target}$ = $\theta$\;
\For{episode = 1, \dots, M}{
    Initialize the environment and receive initial state $s_t$\;
    \For{$t = 1, \dots, N_{\mathrm{BS}}$}{
    with probability $\epsilon$ select random action $a_t$\;
    otherwise select $a_t = \arg\max_a Q(s_t, a; \theta)$\;
    observe $r_t$ and new state $s_{t+1}$\;
    store transition $\{s_t,a_t,r_t,s_{t+1}\}$ in $\mathcal{D}$\;
    \If{memory is full}{
    Sample random mini-batch of transitions $\{s_t,a_t,r_t,s_{t+1}\}$ from $\mathcal{D}$\;
    For each sample $(s_i, a_i, r_i, s_{i+1})$, set:\;
    Set $y_i$ = $r_i$ if episode terminates at step $i + 1$\;
    otherwise set $r_i$ + $\gamma \max_{a'} Q(s_{i+1}, a'$; $\theta_{\mathrm{target}}$)\;
    adopt the stochastic gradient descent (SGD) to update the weights $\theta$ by training DQN according to eq. ~\eqref{loss}\;
    perform $\theta_{target} = \theta$ every $\tau$ steps\;

        }
    }
    }
\caption{DQN algorithm}

\end{algorithm}

The training procedure for the proposed DQN is outlined in Algorithm 1. Initially, the algorithm sets up the necessary parameters (Lines 1$\sim$3) and then iterates over $M$ episodes. Each episode consists of $T$ steps and begins with resetting the environment and states (Line 5). Action selection is performed using the $\epsilon$ greedy policy (Lines 7–8), which balances exploration of new actions with exploitation of past experience. Upon executing an action $a_t$, the agent receives a reward $r_t$ and transitions to a new state $s_{t+1}$. This transition tuple $(s_t, a_t, r_t, s_{t+1})$ is stored in the experience replay memory $D$, which operates in a first-in-first-out manner (Line 10). Once the memory $D$ contains more entries than the predefined minibatch size, network training starts (Lines 11$\sim$15). A random minibatch of transition tuples is sampled from $D$ to update the DQN using past experiences. The target network provides the target value $y_t$ (Line 13), which is used to evaluate the actions chosen by the primary $Q$ network. The loss function, given in \eqref{loss}, guides the update of the main $Q$-network parameters $\theta$ (Line 14), while the target network parameters $\theta_{\text{target}}$ are periodically synchronized every $\tau$ training steps (Line 15). For the single-BS scenario, each episode concludes after one step, as the environment and states remain static. In the multi BS scenario, BSs are deployed incrementally over successive time steps. In this case, each episode deploys one BS per step, and the environment is updated after each deployment to reflect changes in network conditions, such as coverage and EMF exposure. The DQN agent adapts its strategy based on real time feedback, progressively refining placement decisions. This setup mirrors real world scenarios, where network densification occurs gradually, and the objective is to optimize the locations of newly added BSs while keeping previously deployed BSs fixed. Upon completion of training, the DQN has learned an effective ABS placement strategy. During deployment, the trained DQN observes the environment state $s_t$ at each step and selects an action that maximizes $r_t$. This process is repeated multiple times until the optimal BS locations are identified. Upon executing action $a_t$, the environment provides a reward $r_t$ and transitions to a new state $s_{t+1}$, reflecting the updated network configuration and performance after the placement of the new BS. 

\section{Training Setup}\label{training_setup}

To assess the effectiveness of the proposed method, Section \ref{training_setup} describes the training configuration and dataset generation details, and Section \ref{evaluations} presents the corresponding evaluation results.

\subsection{dataset generation}\label{dataproc}
The study focuses on a selected RoI a square area of side 0.8 $km$ within the city of Lyon, specifically encompassing the INSA Lyon campus. Data on 4G and 5G base station (BS) locations within this RoI were obtained from the open source platform www.cartoradio.fr, which provides detailed CSV information including BS position, antenna orientation, height, azimuth, operating frequency, and technology type. Reference coverage and exposure maps were generated using an open source ray tracing simulator \cite{hoydis2023sionnartdifferentiableray, egea2019vehicular}, incorporating building footprints from OpenStreetMap. Exposure calculations followed methodologies from our previous work and the literature \cite{egea2019vehicular, egea2023generation}, ensuring realistic network configurations by applying the BS parameters. For model training, the dataset was split into training and validation sets, with a validation ratio of 0.1 used for testing purposes. The hyperparameters of the cGAN model, such as the number of layers, kernel size, and activation function, were selected based on prior work \cite{mallik2022towards} and by empirical validation to balance performance and computational efficiency. The dataset comprises 5k images, and model training was conducted on an Intel Xeon E5 with NVIDIA Tesla V100 32GB machine on Grid'5000 HPC. 
The parameters of the simulations are given in Table \ref{dataparam}.
\begin{table}[t]
\centering
\begin{tabular}{ll} 
\toprule
\textbf{Parameter} & \textbf{Value} \\ 
\midrule
Map Scale         & $800 \times 800$ $m^{2}$ \\
Terrain           & \checkmark \\
Building          & \checkmark \\
Roads             & \checkmark \\
Carrier Frequency & 3.50 GHz  \\
Transmit Power    & 0 dBm \\
TX type           & Isotropic (Vertical) \\
RX type           & half-wave dipole \\
TX dimension    & $8 \times 4$ (rows × columns)
 \\
\bottomrule
\end{tabular}
\caption{Parameters of the INSA dataset.}
\label{dataparam}
\end{table}
\subsubsection{Data Structure}
In Auto−RSS \cite{auto-rss}, the input-output image has dimensions of $N \times N \times 1$. The BS location was used as the input as a binary image and the output was the coverage image. In this work, following the system model in \cite{auto-rss}, we enhance this input and output data into $N \times N \times 2$. The input tensor consists of a binary map of buildings and Tx map. The output tensor contains the RSS value, and the EMF exposure of the given environment generated from the simulator (described in \cite{zugno2020toward,egea2023generation, hoydis2023sionnartdifferentiableray}). To handle this data, the cGAN architecture is described in Sec. \ref{archi} and trained to predict RSS and exposure simultaneously. 

\subsubsection{Data processing}
\textbf{Data conversion}
To generate the RSS and EMF exposure map, the received power $P_RX$ (in $W$) data was converted into grayscale images using Min-Max normalization, with the minimum value of 0 W and the maximum value of 4.77e-10 W. The data was scaled to dBm. While the upper limit is higher than physically reasonable, this pair of values was chosen fixed to enable convenient gray-level pixel step allocation. It should be noted that modifying the step resolution, either by reducing or increasing, does not considerably affect prediction performance. To measure the exposure data at a given pixel, we have followed our previous work in \cite{egea2019vehicular,egea2023generation}. Based on our work in \cite{mallik2022towards, auto-rss} the 
environment was converted also into grayscale images where buildings corresponds to pixel value 0 and no building areas to pixel value 1. The parameters of the scenario to generate the dataset is given in Table \ref{dataparam}. To fill the missing part of the RSS and exposure maps, we utilize nearest neighbor interpolation, which approximates the missing value with a weighted sum of the pixel values of the adjacent
locations. \\
\textbf{Augmentation}
It is well known that the performance of NN training generally improves as the dataset size increases; in other words, more data often leads to more accurate outcomes. To enhance the dataset, augmentation technique was applied based on flipping operations. Specifically, the entire dataset is flipped vertically and horizontally, to increase the size of the dataset. The building map is also included as an augmented input feature for each dataset sample. The augmented dataset is then partitioned into 90\% for training and 10\% for testing. This ensures a balanced evaluation of the model’s generalization ability. In addition, such augmentation allows the NN to capture diverse propagation patterns and reduces overfitting.
\subsection{NPE-GAN}
The training configuration and hyperparameters for NPE-GAN are given in table below:

\begin{table}[ht]
\centering
\begin{tabular}{lc}
\toprule
\multicolumn{1}{c}{Parameters} & \multicolumn{1}{c}{Value} \\ 
\midrule

Optimizer                        &  $G$ : ADAM, $D$: SGD  \\ 
Kernel Initializer            &  he normal     \\
Non-Linearity                 & $G$: ReLU, $D$: LeakyReLU \\ 
final layer activation        & $G$ : Tanh, $D$: sigmoid\\ 

Loss function  & $G$: MAE, $D$: BCE\\

Epochs & 100 \\ 
\bottomrule

\end{tabular}
\caption{NPE-GAN: Generator and Discriminator training configuration.}

\label{cgan}
\end{table}

\subsection{GAN-DQN}
The environment of the DRL method is the same as NPE-GAN. The RL based method, a CNN model is chosen to be the agent (details of the model architecture is given in \ref{agentt}. The hyperparameter of the GAN-DQN training is given in Table \ref{hypdqn}:
\begin{table}[ht]
\centering
\begin{tabular}{ll} 
\toprule
\textbf{GAN-DQN training configuration} &  \\ 
\midrule
DRL Algorithm        & DQN \cite{mnih2015human} \\
Reward Function           & Coverage \\
Learning rate          & $1.0 \times 10^{-4}$ \\
Gamma             & 0.95 \\
Mini Batch size & 32  \\
CNN loss & MSE\\
CNN optimizer & Adam\\
Activation & ReLU\\
Number of Episodes & 1000\\

\bottomrule
\end{tabular}
\caption{Hyperparameters of GAN-DQN training.}
\label{hypdqn}
\end{table}

\subsection{Evaluation metrics and baselines}
To evaluate the performance of our systems by root mean square error (RMSE) and mean absolute error (MAE), given by:
\begin{equation}
\text{RMSE} = \sqrt{\frac{1}{n} \sum_{i=1}^{n} \left(\Upsilon_i - {\Upsilon'}_i\right)^2}
\end{equation}

\begin{equation}
\text{MAE} = {\frac{1}{n} \sum_{i=1}^{n} \left(\Upsilon_i - {\Upsilon'}_i\right)}
\end{equation}
where $(\Upsilon_i - {\Upsilon'}_i)$ denotes the error between the reference coverage $\Upsilon_i$ and the predicted coverage ${\Upsilon'_i}$ and $n$ is the number of samples in the dataset.
The structural similarity index (SSIM) captures the observed change in the structural information of the picture. The SSIM index is calculated on various windows of an image. The measure between two windows ${x}$ and ${y}$ of common size ${ N\times N}$ is given below:

\begin{equation}
  \textrm{SSIM(x,y)} =\frac{(2\mu_x\mu_y +c_1)(2\sigma_{xy}+ c_2)}{(\mu^2_x+\mu^2_y+ c_1)(\sigma^2_x + \sigma^2_y + c_2)}  
\end{equation}

where, $\mu _{x}$ and $\mu _{y}$ are the pixel sample mean of $x$ and $y$, $\mu _{y}$. $\sigma _{x}^{2}$ and $\sigma _{y}^{2}$ are the variance of ${x}$ and ${y}$, $\sigma_{xy}$ is the covariance of ${x}$ and ${y}$. ${ c_{1}=(k_{1}L)^{2}}$ and ${ c_{2}=(k_{2}L)^{2}}$ are the variables to stabilize the division with weak denominator, where ${L}$ is the dynamic range of the pixel-values. ${ k_{1}}$ and ${ k_{2}}$ are set to $0.01$ and $0.03$.
\subsubsection*{Baselines for NPE-GAN comparison:}

\textbf{RadioUnet} \cite{levie2020radiounetfastradiomap} is an ML-based network performance estimation method that
extends the UNet architecture by employing two UNets \cite{ronneberger2021u}. Each UNet comprises 8 encoder layers with convolution, ReLU, and Maxpooling layers, followed by 8 decoder layers with transposed convolution and ReLU layers. The encoders and decoders are concatenated, as in the original UNet architecture. RadioUNet uses curriculum training to enhance training. In the first stage, the first UNet is trained for a specific number of epochs, with the second UNet frozen. In the second stage, the second UNet is trained using the features and the output of the first UNet.\\

\begin{figure*}[t]
    \centering
    \includegraphics[scale=0.6]{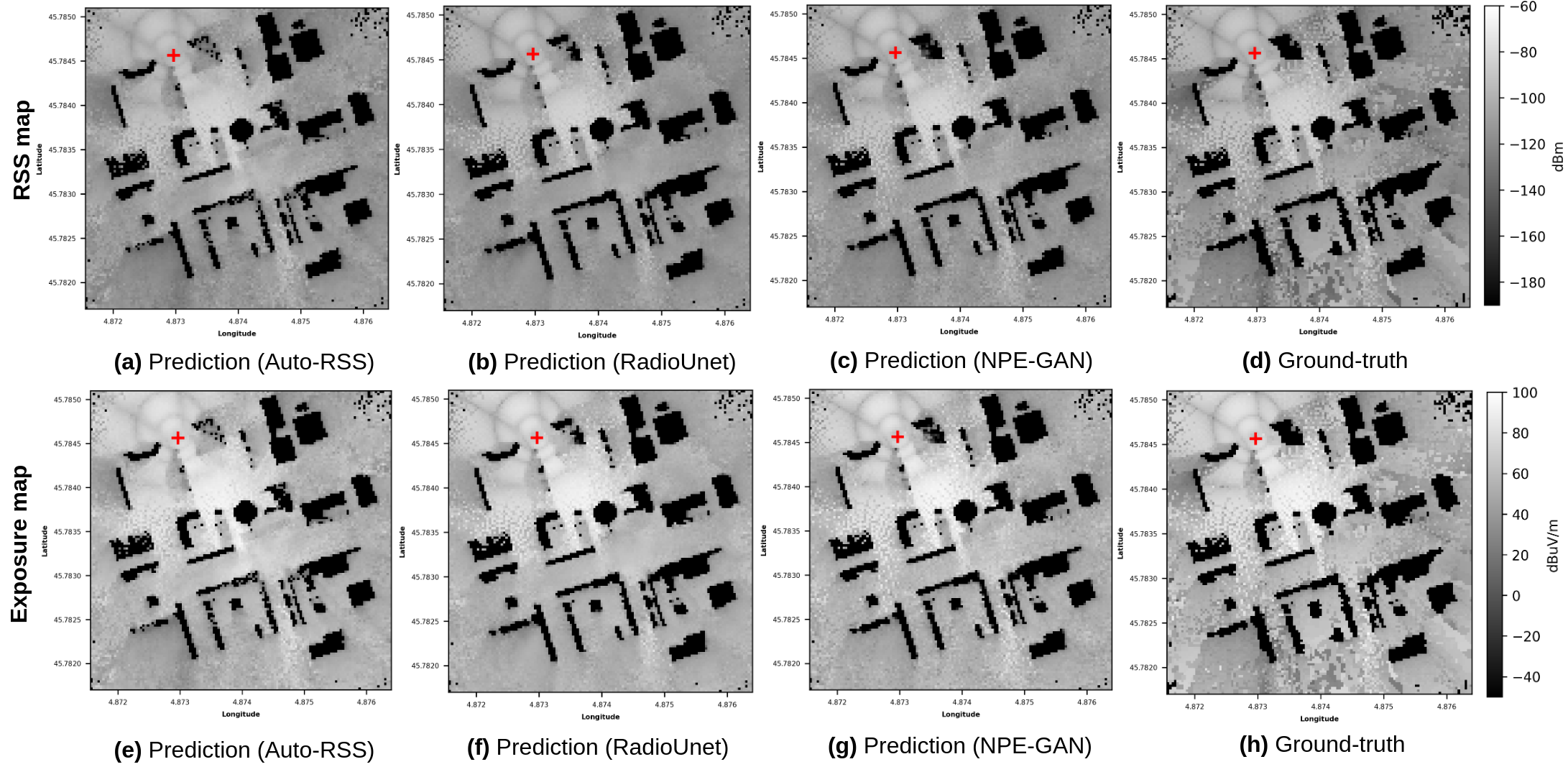}
    \caption{Experimental results for NPE-GAN: 1st row: RSS estimations, 2nd row: EMF Exposure estimation, red cross reveals
the transmitter locations. (d) and (h) are the reference maps from the simulator for comparison.}
    \label{npe_ganres}
\end{figure*}
\textbf{Auto-RSS} \cite{auto-rss}, our previous work which encompasses a CNN based autoencoder. The CNN has four encoder and decoder layers with a bottleneck layer. The input and output dimension are matched to the proposed method.

\subsubsection*{Baseline methods for GAN-DQN comparison:}

\textbf{Random search:} In this baseline strategy, BSs are distributed uniformly at random across the deployable candidate $\mathcal{B}$, without any form of learning or optimization. This purely stochastic placement acts as a performance lower bound because it neglects all site specific characteristics or environmental factors. Despite its simplicity, such random spatial deployment is widely used as a modeling assumption in stochastic geometry analyses \cite{haenggi2013stochastic}. While clearly suboptimal, it highlights the potential gains achievable when informed placement decisions are incorporated.

\textbf{Brute Force:} The goal of a brute force (BF) algorithm is to find the configuration that maximizes the target objective by exploring the entire solution space. For the single-BS scenario on a $128 \times 128$ grid, this search considers all possible positions, guaranteeing the optimal placement. However, as the number of BSs and the map size increase, the complexity grows exponentially, making the method computationally infeasible for multi BS deployments. To maintain practicality, our experiments restrict the exhaustive search to 3k representative locations instead of the full combinatorial space, serving as the multi BS baseline. This reduction allows us to evaluate a manageable subset of configurations while still capturing diverse placement scenarios. This approach optimizes the placement of multiple BSs simultaneously, while BF provides a strong reference for comparison, but at a higher computational cost.

\section{Evaluation}\label{evaluations}
This section provides a performance analysis of the NPE-GAN model (Section \ref{npe_gan_perf}) and the GAN-DQN framework (Section \ref{gan_dqn_perf}), supported by both qualitative and quantitative evaluations.

\subsection{NPE-GAN Performance}\label{npe_gan_perf}
The estimated RSS and EMF exposure estimated by the proposed NPE-GAN are presented in Fig. \ref{npe_ganres}, together with comparative results obtained from RadioUnet and Auto-RSS. The predicted maps produced by NPE-GAN demonstrate a remarkable alignment with the corresponding reference, as shown in Fig. \ref{npe_ganres}-d and in Fig. \ref{npe_ganres}-h for the RSS and EMF exposure, respectively. Comparing the performance of the baseline models, it can be observed that RadioUnet achieves a marginal improvement over Auto-RSS. Both models consistently struggle in accurately capturing the features at the edge regions of the RoI, where signal variations are typically more complex. Similar performance trend is observed in the exposure map predictions displayed in the second row of Fig. \ref{npe_ganres}. 
Unlike the baselines, NPE-GAN uses an adversarial discriminator that explicitly penalizes unrealistic boundary structures, enabling the generator to better reconstruct the steep gradients and irregular propagation patterns often found at the edges of the RoI.
This adversarial feedback forces the generator to learn higher frequency spatial details, which standard regression-based models tend to smooth out, explaining its superior performance in the edge zones seen in Fig. \ref{npe_ganres}.
As a result, NPE-GAN produces sharper transitions and more representations of shadowing effects near obstacles and region boundaries or edges, where traditional CNN models typically lose accuracy.

To further validate the proposed method, Table \ref{feature} presents the accuracy of the estimation impact of data augmentation and different size of filters in the cGAN model. Horizontal and vertical flips are used to enhance the dataset for training the model. This exposes the model to a wider range of input-output patterns, thus reducing overfitting. This helps the model to be robust in terms of noise and variability, as shown in Table \ref{feature}, increasing the filters lowers the MAE error below 5 dB. Whereas without augmentation and a lower number of filters in the layers, such as 32, RMSE is at 17.13 dB.

This performance is confirmed in Table \ref{rmse}, where NPE-GAN achieves the lowest error values with a RMSE of approximately 7.3 dB, an MAE of roughly 4 dB, and an SSIM around 0.87 for RSS. 
\begin{table}[t]
\centering
\begin{tabular}{cccccc} \toprule

Parameter & Filters & MAE & RMSE & SSIM\\
\midrule
\multirow{3}{*}{\rotatebox{90}{RSS}} 

w/o aug  & 32 & 10.57 & 17.13 & 0.40\\ 
aug  & 32 & 9.64 & 16.14 & 0.43\\
aug  & 128 & 7.27 & 10.77 & 0.77\\
aug  & \textbf{256} & \textbf{4.41} & \textbf{7.30} & \textbf{0.87}\\
\midrule
\multirow{3}{*}{\rotatebox{90}{exposure}} 
w/o aug  & 32 & 10.57 & 17.13 & 0.40\\ 
aug  & 32 & 9.64 & 16.14 & 0.43\\
aug  & 128 & 7.27 & 10.77 & 0.77\\
aug  & \textbf{256}  & \textbf{3.72} & \textbf{6.33}  & \textbf{0.85} \\
\bottomrule
\end{tabular}
\caption{RMSE, MAE and SSIM of the estimated RSS and exposure by our approach with or without augmentation and increasing number of filters.}
\label{feature}
\end{table}

\begin{table}[ht]
\centering
\begin{tabular}{cccccc} \toprule
 & Model & MAE & RMSE & SSIM \\ 
\midrule\\
\multirow{3}{*}{\rotatebox{90}{RSS}} 
  & Auto-RSS   & 7.37 & 15.29 & 0.76 \\
  & RadioUnet  & 4.53 & 8.04  & 0.83 \\
  & \textbf{NPE-GAN} & \textbf{4.41} & \textbf{7.30} & \textbf{0.87} \\\\
\midrule\\
\multirow{3}{*}{\rotatebox{90}{Exposure}} 
  & Auto-RSS   & 6.45 & 13.70  & 0.67 \\
  & RadioUnet  & 3.89 & 7.05  & 0.80 \\
  & \textbf{NPE-GAN}    & \textbf{3.72} & \textbf{6.33}  & \textbf{0.85} \\\\
\bottomrule
\end{tabular}
\caption{RMSE of the estimated exposure values using our approach.}
\label{rmse}
\end{table}
The error comparison of the baseline models with NPE-GAN is provided in Table \ref{rmse}. The Auto-RSS model achieves a MAE of 7.37 dB for RSS estimation and 6.45 dB for exposure estimation, with SSIM scores of 0.76 and 0.67, respectively. In comparison, RadioUnet demonstrates improved performance over Auto-RSS, yielding MAE values of 4.53 dB for RSS and 3.89 dB for EMF exposure, while maintaining a SSIM of 0.83 and 0.80 in both cases. In contrast, the proposed NPE-GAN substantially outperforms both baselines, achieving MAE of 4.41 dB for RSS estimation and 3.72 dB for EMF exposure estimation, with consistently higher SSIM values. These results demonstrate the advantage of the generator–discriminator design of NPE-GAN, which enables accurate and simultaneous estimation of multiple network performance metrics.

\subsection{GAN-DQN Performance Comparison}\label{gan_dqn_perf}

Table~\ref{comp} compares the deployment strategies of Random Search, Brute-Force, and the proposed GAN-DQN for both single and 2 BS scenarios under EMF level of 6 $V/m$ which the safety limit provided by ICNIRP and WHO. The Brute Force approach, which serves as the global optimal benchmark, achieves the highest coverage with $64.66\%$ for a single BS and $75.05\%$ for two BSs. In contrast, Random Search yields considerably lower coverage, only $45.80\%$ (1 BS) and $56.55\%$ (2 BS). GAN-DQN substantially improves to a near-optimal solution, achieving $59.57\%$ coverage with one BS and $67.15\%$ with two BSs. These results demonstrate that GAN-DQN narrows the gap with the optimal Brute-Force method, validating the effectiveness of agent guided BS placement. Importantly, in all deployment cases, the $ER$ observed during training and evaluation was consistently above the reward threshold, reflecting that the algorithm actively penalizes high EMF exposure. The relatively small difference in coverage between GAN-DQN and Brute-Force indicates that GAN-DQN is capable of delivering near-optimal performance while enforcing exposure constraints through its reward design based on RSS and EMF exposure estimates from NPE-GAN.
\begin{table}[t]
\centering
\begin{tabular}{cccc} \toprule
Model & Coverage - 1 BS & Coverage - 2BS\\
 
\midrule
Brute-Force  & 64.66\% & 75.05\%\\ 
Random Search  & 45.80\% & 56.55\%\\
GAN-DQN  & 59.57\% & 67.15\% \\
\bottomrule

\end{tabular}
\caption{Comparison of single and multi BS deployment between GAN-DQN, BF, and random search.}
\label{comp}
\end{table}
\begin{figure}[ht]
    \centering
    \includegraphics[scale=0.4]{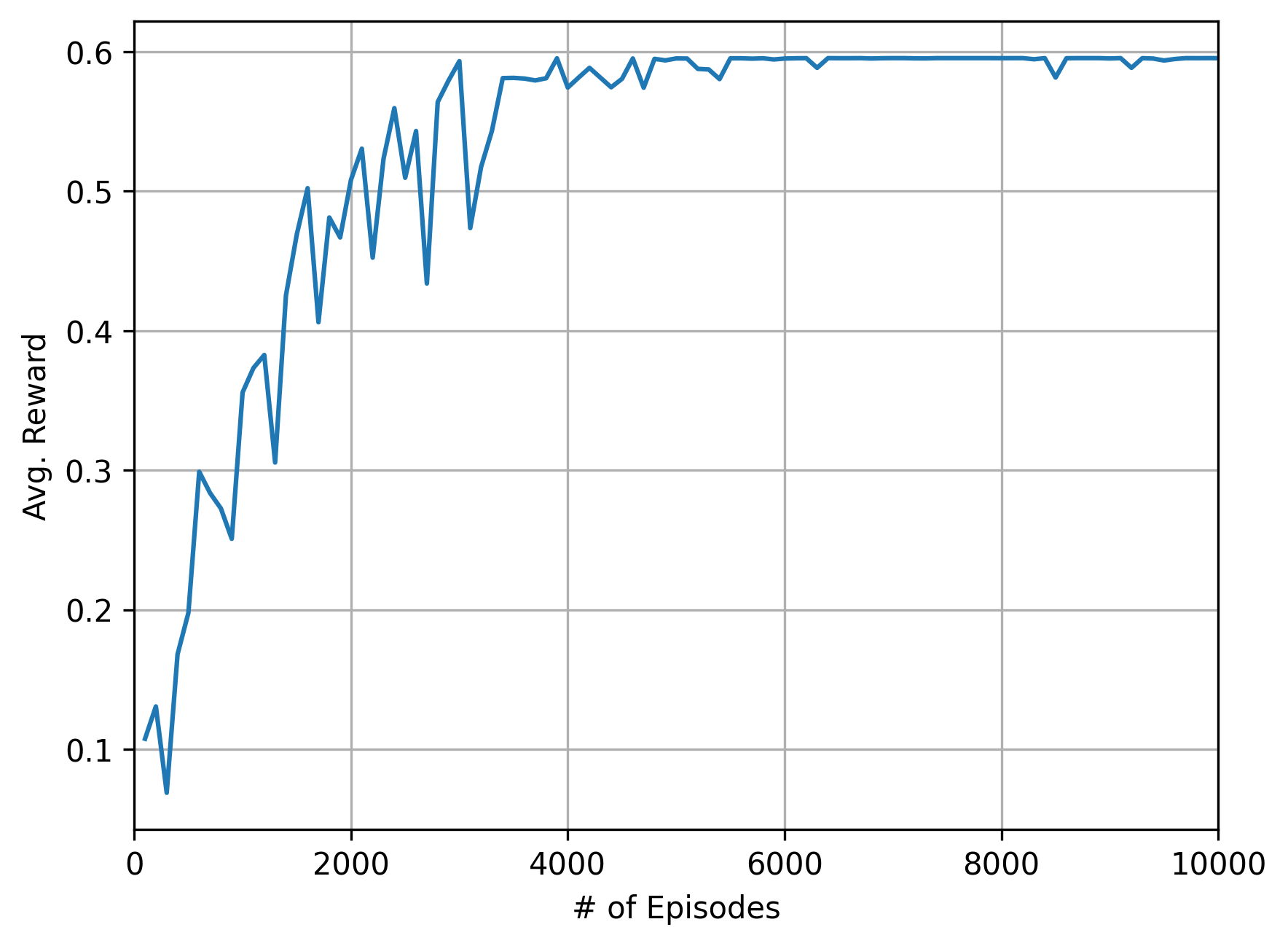}
    \caption{Convergence behavior for 1 BS deployment GAN-DQN over episodes.}
    \label{converge}
\end{figure}

The training convergence for the single-BS deployment scenario is illustrated in Fig. \ref{converge}. GAN-DQN’s performance improves steadily and stabilizes after approximately 5k training episodes.
GAN-DQN achieves near optimal performance efficiently, demonstrating effective learning and stable convergence.

\begin{figure*}[t]
    \centering
    \includegraphics[scale=0.7]{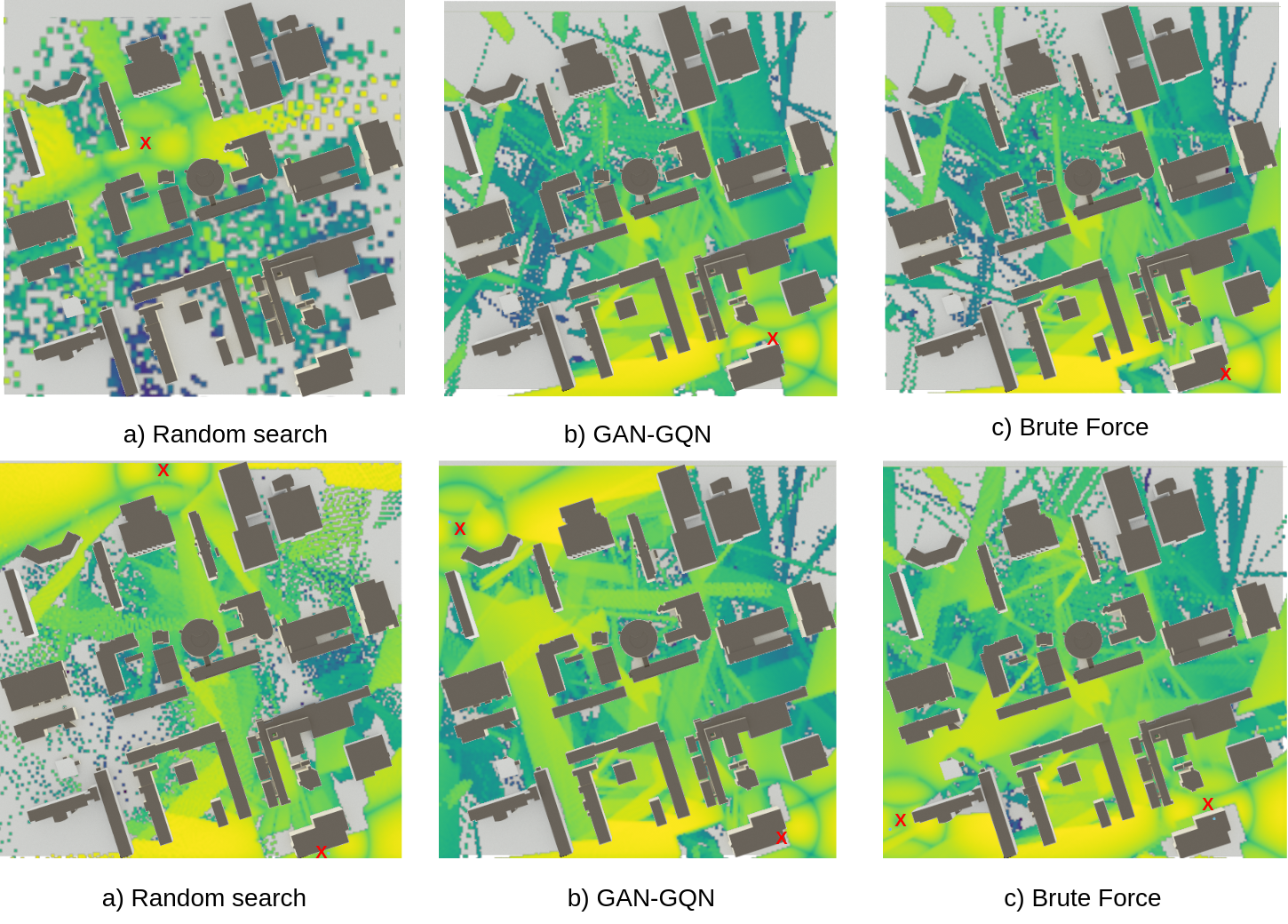}
    \caption{GAN-DQN 1 and 2 BS coverage maps, 1st row shows 1 BS and 2nd rows is 2 BS placed and red $X$ reveals the BS location. We use Sionna visualization for nicer 3D looks.}
    \label{con_dqn}
\end{figure*}
\subsubsection*{Coverage}

Fig. \ref{con_dqn} illustrates the coverage performance for both single (Fig \ref{con_dqn} - 1st row) and 2 BS (Fig. \ref{con_dqn}-2nd row) deployments. The Brute Force search method exhaustively evaluates all possible BS placement configurations, yielding the highest achievable coverage; however, this comes at the cost of prohibitive computational complexity. In the random search approach, BSs are distributed randomly throughout the deployment area, resulting in suboptimal coverage and gaps. This shows the limitation of the non optimized deployment.  The proposed GAN-DQN, by integrating the DQN algorithm with rapid and accurate network performance predictions from NPE-GAN, efficiently identifies near optimal BS placements with EMF constraints. This approach achieves coverage levels comparable to BF search while substantially reducing computation time. Moreover, the DQN component enables the agent to learn long term placement strategies by iteratively updating the Q values based on the expected reward, effectively balancing exploration and exploitation during training. The combination of GAN and DQN thus provides a scalable and adaptive solution for large scale BS deployment optimization.

\subsubsection{Time efficiency analysis}
\begin{table}[t]
\centering
\begin{tabular}{cccc} \toprule
Model & 1 BS & 2 BS\\
 
\midrule
Brute-Force  & 181 & 5490\\ 
Gan-DQN  & 5 & 8 \\
\bottomrule

\end{tabular}
\caption{Time comparison of single and multi BS deployment between GAN-DQN and BF (in seconds).}
\label{time}
\end{table}
A comparison of computation times between the proposed GAN-DQN approach and the baseline Brute Force (BF) search is presented in Table \ref{time}. The results shows a significant advantage of GAN-DQN in terms of inference speed, as it is capable of generating deployment decisions within seconds, even in scenarios involving multiple BSs. For a single BS, BF requires 181 seconds, whereas GAN-DQN requires the deployment in only 10 seconds, representing an over 18 times reduction in computation time. The advantage becomes even more visible for two BSs. In 2 BS case, BF’s runtime increases to 5,490 seconds, while GAN-DQN achieves the task in merely 16 seconds on a 16 core server with 512 RAM. These results clearly demonstrate that BF search becomes quickly impractical as the number of BSs grows, whereas GAN-DQN maintains millisecond scale inference even for multi BS scenarios.

The sequential deployment strategy adopted by GAN-DQN further contributes to its scalability, allowing the method to accommodate larger network sizes without a substantial increase in computation time. This characteristic positions GAN-DQN as a highly suitable solution for real-time network optimization. Additionally, the method’s low latency inference enables rapid exploration of alternative deployment strategies. This is particularly beneficial for dynamic network environments and adaptive planning scenarios. Moreover, the proposed approach is fast, scalable, and incorporates EMF exposure constraints, ensuring that all evaluated deployment solutions comply with predefined EMF thresholds.
\section{Discussion}\label{dis}
GAN-DQN approaches the optimal solutions obtained through brute-force search, achieving approximately 90–92\% of the coverage offered by exhaustive placement in both the single and two BS scenarios. The residual performance gap arises from the different search strategies. BF evaluates all candidate locations and is therefore guaranteed to find the global optimum. GAN-DQN explores the action space selectively and approximates the value function, converging to high quality  placements. In practice, however, this small loss in optimality is offset by a gain in computational efficiency. In our experiments, the brute force (BF) method required hours to evaluate a subset of a few thousand placement configurations, far from the entire search space. In contrast, GAN-DQN drastically reduces the computation time and identified near optimal 2 BS placements in under 10 seconds. This makes GAN-DQN a much more practical solution for real time or large scale deployment planning, where the computational cost of exhaustive search quickly becomes impractical.

Beyond performance gains, the method shows robust scalability, enabling practical deployment in dense urban environments and complex propagation scenarios while considering EMF exposure constraints. The current framework shows limitations that needs further investigation. In particular, evaluating its scalability to very large metropolitan areas with heterogeneous environments remains an open challenge.
Incorporating inter cell dependencies would require a significantly more complex optimization framework. Additionally, extending the system to joint optimization of BS placement and transmit power is another promising research direction.

\section{Conclusion}\label{conclude}

In this work, we introduce an EMF aware reinforcement learning-based framework that uses a cGAN based model for optimal BS deployment in next generation wireless networks. Specifically, our approach, GAN-DQN, combines the DQN algorithm with fast and accurate network performance predictions generated by NPE-GAN. This conditional generative model is designed as a digital twins (DT) capable of predicting RSS and EMF exposure rapidly. By integrating these components, GAN-DQN effectively learns deployment policies that are adaptive to site specific channel characteristics while simultaneously optimizing coverage following EMF exposure safety limits. Moreover, the underlying GAN-DQN framework is highly adaptable and can be extended to a wide range of network optimization tasks within a DT context. By redefining the state representation and reward function, GAN-DQN can address challenges such as mobility management, adaptive link control, or energy efficiency of a mobile network. Its flexibility further allows application to alternative deployment scenarios, for energy-efficient network design. 

\bibliographystyle{ieeetr}

\bibliography{biblio.bib}

@article{mnih2015human,
  title={Human-level control through deep reinforcement learning},
  author={Mnih, Volodymyr and Kavukcuoglu, Koray and Silver, David and Rusu, Andrei A and Veness, Joel and Bellemare, Marc G and Graves, Alex and Riedmiller, Martin and Fidjeland, Andreas K and Ostrovski, Georg and others},
  journal={nature},
  volume={518},
  number={7540},
  pages={529--533},
  year={2015},
  publisher={Nature Publishing Group}
}

@book{haenggi2013stochastic,
  title={Stochastic geometry for wireless networks},
  author={Haenggi, Martin},
  year={2013},
  publisher={Cambridge University Press}
}

@ARTICLE{8788518,
  author={Liu, Yaxi and Huangfu, Wei and Zhang, Haijun and Wang, Haobin and An, Wei and Long, Keping},
  journal={IEEE Access}, 
  title={An Efficient Geometry-Induced Genetic Algorithm for Base Station Placement in Cellular Networks}, 
  year={2019},
  volume={7},
  number={},
  pages={108604-108616},
  doi={10.1109/ACCESS.2019.2933284}}

@ARTICLE{9079581,
  author={Dai, Lingcheng and Zhang, Hongtao},
  journal={IEEE Access}, 
  title={Propagation-Model-Free Base Station Deployment for Mobile Networks: Integrating Machine Learning and Heuristic Methods}, 
  year={2020},
  volume={8},
  number={},
  pages={83375-83386},
  doi={10.1109/ACCESS.2020.2990631}}

@ARTICLE{7161398,
  author={Li, Xu and Guo, Dongning and Grosspietsch, John and Yin, Huarui and Wei, Guo},
  journal={IEEE Transactions on Vehicular Technology}, 
  title={Maximizing Mobile Coverage via Optimal Deployment of Base Stations and Relays}, 
  year={2016},
  volume={65},
  number={7},
  pages={5060-5072},
  doi={10.1109/TVT.2015.2458015}}

@ARTICLE{8119520,
  author={Chatterjee, Shubhajeet and Abdel-Rahman, Mohammad J. and MacKenzie, Allen B.},
  journal={IEEE Wireless Communications Letters}, 
  title={Optimal Base Station Deployment with Downlink Rate Coverage Probability Constraint}, 
  year={2018},
  volume={7},
  number={3},
  pages={340-343},
  doi={10.1109/LWC.2017.2776929}}

@INPROCEEDINGS{8292392,
  author={Prasad, Ganesh and Mishra, Deepak and Hossain, Ashraf},
  booktitle={2017 IEEE 28th Annual International Symposium on Personal, Indoor, and Mobile Radio Communications (PIMRC)}, 
  title={Coverage-constrained base station deployment and power allocation for operational cost minimization}, 
  year={2017},
  volume={},
  number={},
  pages={1-5},
  doi={10.1109/PIMRC.2017.8292392}}

@INPROCEEDINGS{8717805,
  author={Wu, Jiahui and Yu, Peng and Feng, Lei and Zhou, Fanqin and Li, Wenjing and Qiu, Xuesong},
  booktitle={2019 IFIP/IEEE Symposium on Integrated Network and Service Management (IM)}, 
  title={3D Aerial Base Station Position Planning based on Deep Q-Network for Capacity Enhancement}, 
  year={2019},
  volume={},
  number={},
  pages={482-487},
  doi={}}

@article{aaa,
title = "Deep-Reinforcement-Learning-Based Drone Base Station Deployment for Wireless Communication Services",
author = "Tarekegn, \{Getaneh Berie\} and Juang, \{Rong Terng\} and Lin, \{Hsin Piao\} and Munaye, \{Yirga Yayeh\} and Wang, \{Li Chun\} and Bitew, \{Mekuanint Agegnehu\}",
note = "Publisher Copyright: {\textcopyright} 2014 IEEE.",
year = "2022",
month = nov,
day = "1",
doi = "10.1109/JIOT.2022.3182633",
language = "English",
volume = "9",
pages = "21899--21915",
journal = "IEEE Internet of Things Journal",
issn = "2327-4662",
publisher = "Institute of Electrical and Electronics Engineers Inc.",
number = "21",
}

@Article{electronics10232953,
AUTHOR = {Gopi, Sudheesh Puthenveettil and Magarini, Maurizio},
TITLE = {Reinforcement Learning Aided UAV Base Station Location Optimization for Rate Maximization},
JOURNAL = {Electronics},
VOLUME = {10},
YEAR = {2021},
NUMBER = {23},
ARTICLE-NUMBER = {2953},
URL = {https://www.mdpi.com/2079-9292/10/23/2953},
ISSN = {2079-9292},
DOI = {10.3390/electronics10232953}
}

@article{samal,
author = {Samal, Soumya Ranjan and Swain, Kaliprasanna and Bandopadhaya, Shuvabrata and Dandanov, Nikolay and Poulkov, Vladimir and Routray, Sidheswar and Palai, Gopinath},
title = {Dynamic Coverage Optimization for 5G Ultra-dense Cellular Networks Based on Their User Densities},
year = {2022},
issue_date = {Jan 2023},
publisher = {Kluwer Academic Publishers},
address = {USA},
volume = {128},
number = {1},
issn = {0929-6212},
url = {https://doi.org/10.1007/s11277-022-09969-4},
doi = {10.1007/s11277-022-09969-4},

journal = {Wirel. Pers. Commun.},
month = sep,
pages = {605–620},
numpages = {16}
}

@article{interdonato2019downlink,
  title={Downlink training in cell-free massive MIMO: A blessing in disguise},
  author={Interdonato, Giovanni and Ngo, Hien Quoc and Frenger, P{\aa}l and Larsson, Erik G},
  journal={IEEE Transactions on Wireless Communications},
  volume={18},
  number={11},
  pages={5153--5169},
  year={2019},
  publisher={IEEE}
}

@article{SKOCAJ2022403,
title = {Cellular Network Capacity and Coverage Enhancement with MDT Data and Deep Reinforcement Learning},
journal = {Computer Communications},
volume = {195},
pages = {403-415},
year = {2022},
issn = {0140-3664},
doi = {https://doi.org/10.1016/j.comcom.2022.09.005},
url = {https://www.sciencedirect.com/science/article/pii/S0140366422003449},
author = {Marco Skocaj and Lorenzo M. Amorosa and Giorgio Ghinamo and Giuliano Muratore and Davide Micheli and Flavio Zabini and Roberto Verdone},


}

@inproceedings{gong,
author = {Gong, Jiahui and Yu, Qiaohong and Li, Tong and Liu, Haoqiang and Zhang, Jun and Fan, Hangyu and Jin, Depeng and Li, Yong},
title = {Demo: Scalable Digital Twin System for Mobile Networks with Generative AI},
year = {2023},
isbn = {9798400701108},
publisher = {Association for Computing Machinery},
address = {New York, NY, USA},
url = {https://doi.org/10.1145/3581791.3597297},
doi = {10.1145/3581791.3597297},

booktitle = {Proceedings of the 21st Annual International Conference on Mobile Systems, Applications and Services},
pages = {610–611},
numpages = {2},
location = {Helsinki, Finland},
series = {MobiSys '23}
}

@article{article,
author = {Xiao, Zhu and Li, Tong and Ding, Wei and Wang, Dong and Zhang, Jie},
year = {2016},
month = {02},
pages = {n/a-n/a},
title = {Dynamic PCI allocation on avoiding handover confusion via cell status prediction in LTE heterogeneous small cell networks},
volume = {16},
journal = {Wireless Communications and Mobile Computing},
doi = {10.1002/wcm.2662}
}

@ARTICLE{9992177,
  author={Iacoboaiea, Ovidiu and Krolikowski, Jonatan and Houidi, Zied Ben and Rossi, Dario},
  journal={IEEE Communications Magazine}, 
  title={From Design to Deployment of Zero Touch Deep Reinforcement Learning WLANs}, 
  year={2023},
  volume={61},
  number={2},
  pages={104-109},
  doi={10.1109/MCOM.002.2200318}}

@ARTICLE{8316776,
  author={Sekander, Silvia and Tabassum, Hina and Hossain, Ekram},
  journal={IEEE Communications Magazine}, 
  title={Multi-Tier Drone Architecture for 5G/B5G Cellular Networks: Challenges, Trends, and Prospects}, 
  year={2018},
  volume={56},
  number={3},
  pages={96-103},
  doi={10.1109/MCOM.2018.1700666}}

@INPROCEEDINGS{8658363,
  author={Cicek, Cihan Tugrul and Gultekin, Hakan and Tavli, Bulent and Yanikomeroglu, Halim},
  booktitle={2019 1st International Conference on Unmanned Vehicle Systems-Oman (UVS)}, 
  title={UAV Base Station Location Optimization for Next Generation Wireless Networks: Overview and Future Research Directions}, 
  year={2019},
  volume={},
  number={},
  pages={1-6},
  doi={10.1109/UVS.2019.8658363}}

@INPROCEEDINGS{10225848,
  author={Sun, Li and Hou, Jing and Chapman, Richard},
  booktitle={IEEE INFOCOM 2023 - IEEE Conference on Computer Communications Workshops (INFOCOM WKSHPS)}, 
  title={Multi-Agent Deep Reinforcement Learning for Access Point Activation Strategy in Cell-Free Massive MIMO Networks}, 
  year={2023},
  volume={},
  number={},
  pages={1-6},
  doi={10.1109/INFOCOMWKSHPS57453.2023.10225848}}

@article{arora2003approximation,
  title={Approximation schemes for NP-hard geometric optimization problems: A survey},
  author={Arora, Sanjeev},
  journal={Mathematical Programming},
  volume={97},
  number={1},
  pages={43--69},
  year={2003},
  publisher={Springer}
}

@article{yoon2025study,
  title={Study of Optimal Base Station Deployment for UAM Operations in an Urban Environment Based on a Genetic Algorithm},
  author={Yoon, Minsang and Park, Jiseok and Park, Bosung and Jin, Taekyeong and Choo, Hosung},
  journal={IEEE Access},
  year={2025},
  publisher={IEEE}
}

@article{LI2025111431,
title = {Optimization of 5G base station deployment based on quantum genetic algorithm in outdoor 3D map},
journal = {Computer Networks},
volume = {269},
pages = {111431},
year = {2025},
issn = {1389-1286},
doi = {https://doi.org/10.1016/j.comnet.2025.111431},
url = {https://www.sciencedirect.com/science/article/pii/S1389128625003986},
author = {Jianpo Li and Jinjian Pang and Binfeng Jiang and Qi Xu and Enyuan Zhang},
}

@article{LIAQ2025103983,
title = {Utilization of machine learning in future wireless networks for resource optimization: A survey},
journal = {Ad Hoc Networks},
volume = {178},
pages = {103983},
year = {2025},
issn = {1570-8705},
doi = {https://doi.org/10.1016/j.adhoc.2025.103983},
url = {https://www.sciencedirect.com/science/article/pii/S1570870525002318},
author = {Mudassar Liaq and Sana Sharif and Sherali Zeadally and Waleed Ejaz},

}

@article{LI2023119224,
title = {6G shared base station planning using an evolutionary bi-level multi-objective optimization algorithm},
journal = {Information Sciences},
volume = {642},
pages = {119224},
year = {2023},
issn = {0020-0255},
doi = {https://doi.org/10.1016/j.ins.2023.119224},
url = {https://www.sciencedirect.com/science/article/pii/S0020025523008095},
author = {Kuntao Li and Weizhong Wang and Hai-Lin Liu}
}

@inproceedings{chouvardas2016method,
  title={A method to reconstruct coverage loss maps based on matrix completion and adaptive sampling},
  author={Chouvardas, Symeon and Valentin, Stefan and Draief, Moez and Leconte, Mathieu},
  booktitle={2016 IEEE International Conference on Acoustics, Speech and Signal Processing (ICASSP)},
  pages={6390--6394},
  year={2016},
  organization={IEEE}
}

@article{saito2019two,
  title={Two-step path loss prediction by artificial neural network for wireless service area planning},
  author={Saito, Kentaro and Jin, Yongri and Kang, CheChia and Takada, Jun-ichi and Leu, Jenq-Shiou},
  journal={IEICE Communications Express},
  year={2019},
  publisher={The Institute of Electronics, Information and Communication Engineers}
}

@ARTICLE{8794603,
  author={Li, Zhuo and Cao, Jiannong and Wang, Hongwei and Zhao, Miao},
  journal={IEEE Journal on Selected Areas in Communications}, 
  title={Sparsely Self-Supervised Generative Adversarial Nets for Radio Frequency Estimation}, 
  year={2019},
  volume={37},
  number={11},
  pages={2428-2442},
  doi={10.1109/JSAC.2019.2933779}}

@article{mallik2022towards,
  title={Towards Outdoor Electromagnetic Field Exposure Mapping Generation Using Conditional {GAN}s},
  author={Mallik, Mohammed and Tesfay, Angesom Ataklity and Allaert, Benjamin and Kassi, R{\'e}dha and Egea-Lopez, Esteban and Molina-Garcia-Pardo, Jose-Maria and Wiart, Joe and Gaillot, Davy P and Clavier, Laurent},
  journal={Sensors},
  volume={22},
  number={24},
  pages={9643},
  year={2022},
  publisher={MDPI}
}

@inproceedings{mallik2023eme,
  title={{EME-GAN}: A Conditional Generative Adversarial Network based Indoor {EMF} Exposure Map Reconstruction},
  author={Mallik, Mohammed and Allaert, Benjamin and Tesfay, Angesom and Gaillot, Davy P and Wiart, Joe and Clavier, Laurent},
  booktitle={29{\^A} Colloque sur le traitement du signal et des image},
  volume={23},
  pages={745--748},
  year={2023}
}

@article{mirza2014conditional,
  title={Conditional generative adversarial nets},
  author={Mirza, Mehdi and Osindero, Simon},
  journal={arXiv preprint arXiv:1411.1784},
  year={2014}
}

@misc{ronneberger2021u,
  title={{U-Net}: convolutional networks for biomedical image segmentation. ArXiv150504597 Cs. Published online May 18, 2015},
  author={Ronneberger, O and Fischer, P and Brox, T},
  year={2021}
}

@ARTICLE{339880,
  author={Andersen, J.B. and Rappaport, T.S. and Yoshida, S.},
  journal={IEEE Communications Magazine}, 
  title={Propagation measurements and models for wireless communications channels}, 
  year={1995},
  volume={33},
  number={1},
  pages={42-49},
  doi={10.1109/35.339880}}

@article{yun2015ray,
  title={Ray tracing for radio propagation modeling: Principles and applications},
  author={Yun, Zhengqing and Iskander, Magdy F},
  journal={IEEE access},
  volume={3},
  pages={1089--1100},
  year={2015},
  publisher={IEEE}
}

@inproceedings{egea2023generation,
  title={Generation of electromagnetic exposure maps for 5G communications},
  author={Egea-Lopez, Esteban and Mallik, Mohammed and Clavier, Laurent and Gaillot, Davy P},
  booktitle={18th European Conference on Antennas and Propagation (EuCAP2024): Glasgow, del 17 al 22 de marzo de 2024},
  year={2023},
  organization={IEEE}
}

@misc{demir2018patchbasedimageinpaintinggenerative,
      title={Patch-Based Image Inpainting with Generative Adversarial Networks}, 
      author={Ugur Demir and Gozde Unal},
      year={2018},
      eprint={1803.07422},
      archivePrefix={arXiv},
      primaryClass={cs.CV},
      url={https://arxiv.org/abs/1803.07422}, 
}

@misc{levie2020radiounetfastradiomap,
      title={RadioUNet: Fast Radio Map Estimation with Convolutional Neural Networks}, 
      author={Ron Levie and Çağkan Yapar and Gitta Kutyniok and Giuseppe Caire},
      year={2020},
      eprint={1911.09002},
      archivePrefix={arXiv},
      primaryClass={eess.SP},
      url={https://arxiv.org/abs/1911.09002}, 
}

@misc{auto-rss,
author = {Mallik, M. and Villemaud, Guillaume},
year = {2025},
month = {05},
pages = {1-6},
title = {Deep Learning Based Received Signal Strength Estimation},
doi = {10.1109/ICMLCN64995.2025.11140477}
}

@misc{hoydis2023sionnartdifferentiableray,
      title={Sionna RT: Differentiable Ray Tracing for Radio Propagation Modeling}, 
      author={Jakob Hoydis and Fayçal Aït Aoudia and Sebastian Cammerer and Merlin Nimier-David and Nikolaus Binder and Guillermo Marcus and Alexander Keller},
      year={2023},
      eprint={2303.11103},
      archivePrefix={arXiv},
      primaryClass={cs.IT},
      url={https://arxiv.org/abs/2303.11103}, 
}

@article{ZHAO2025109844,
title = {Maximizing coverage in UAV-based emergency communication networks using deep reinforcement learning},
journal = {Signal Processing},
volume = {230},
pages = {109844},
year = {2025},
issn = {0165-1684},
doi = {https://doi.org/10.1016/j.sigpro.2024.109844},
url = {https://www.sciencedirect.com/science/article/pii/S016516842400464X},
author = {Le Zhao and Xiongchao Liu and Tao Shang},

}

@INPROCEEDINGS{9149258,
  author={Qiu, Jin and Lyu, Jiangbin and Fu, Liqun},
  booktitle={ICC 2020 - 2020 IEEE International Conference on Communications (ICC)}, 
  title={Placement Optimization of Aerial Base Stations with Deep Reinforcement Learning}, 
  year={2020},
  volume={},
  number={},
  pages={1-6},
  doi={10.1109/ICC40277.2020.9149258}}

@INPROCEEDINGS{10257851,
  author={Zhang, Jiafeng and Zhu, Linshi and Chen, Haowen},
  booktitle={2023 IEEE International Conference on Image Processing and Computer Applications (ICIPCA)}, 
  title={5G Base Station Siting Scheme Based on Improved Immune Genetic Algorithm}, 
  year={2023},
  volume={},
  number={},
  pages={1107-1112},
  doi={10.1109/ICIPCA59209.2023.10257851}}

@article{wang2021fast,
  title={Fast Construction of the Radio Map Based on the Improved Low-Rank Matrix Completion and Recovery Method for an Indoor Positioning System},
  author={Wang, Zhuang and Zhang, Liye and Kong, Qun and Wang, Kangtao},
  journal={Journal of Sensors},
  volume={2021},
  pages={1--12},
  year={2021},
  publisher={Hindawi Limited}
}

@article{zugno2020toward,
  title={Toward standardization of millimeter-wave vehicle-to-vehicle networks: Open challenges and performance evaluation},
  author={Zugno, Tommaso and Drago, Matteo and Giordani, Marco and Polese, Michele and Zorzi, Michele},
  journal={IEEE Communications Magazine},
  volume={58},
  number={9},
  pages={79--85},
  year={2020},
  publisher={IEEE}
}

@article{wahl2005dominant,
  title={Dominant path prediction model for urban scenarios},
  author={Wahl, Ren{\'e} and W{\"o}lfle, Gerd and Wertz, Philipp and Wildbolz, Pascal and Landstorfer, Friedrich},
  journal={14th IST Mobile and Wireless Communications Summit, Dresden (Germany)},
  year={2005}
}

@article{rizk1997two,
  title={Two-dimensional ray-tracing modeling for propagation prediction in microcellular environments},
  author={Rizk, Karim and Wagen, J-F and Gardiol, Fred},
  journal={IEEE Transactions on Vehicular Technology},
  volume={46},
  number={2},
  pages={508--518},
  year={1997},
  publisher={IEEE}
}

@INPROCEEDINGS{6831690,
  author={MacCartney, George R. and Junhong Zhang and Shuai Nie and Rappaport, Theodore S.},
  booktitle={2013 IEEE Global Communications Conference (GLOBECOM)}, 
  title={Path loss models for {5G} millimeter wave propagation channels in urban microcells}, 
  year={2013},
  volume={},
  number={},
  pages={3948-3953},
  doi={10.1109/GLOCOM.2013.6831690}}

@INPROCEEDINGS{6363950,
  author={Piersanti, Stefano and Annoni, Luca Alfredo and Cassioli, Dajana},
  booktitle={2012 {IEEE} International Conference on Communications {(ICC)}}, 
  title={Millimeter waves channel measurements and path loss models}, 
  year={2012},
  volume={},
  number={},
  pages={4552-4556},
  doi={10.1109/ICC.2012.6363950}
}

@ARTICLE{10715541,
  author={Yao, Ming and Wei, Zhaohui and Li, Kun and Frølund Pedersen, Gert and Zhang, Shuai},
  journal={IEEE Transactions on Antennas and Propagation}, 
  title={Prediction of Electromagnetic Field Exposure at 20–100 GHz for Clothed Human Body Using an Adaptively Reconfigurable Architecture Neural Network With Weight Analysis ({RAWA-NN}) Framework}, 
  year={2024},
  volume={72},
  number={12},
  pages={9286-9300},
  doi={10.1109/TAP.2024.3474913}}

@article{tognola2021use,
  title={Use of Machine Learning for the Estimation of Down and Up {L}ink Field Exposure in Multi-Source Indoor WiFi Scenarios},
  author={Tognola, Gabriella and Plets, David and Chiaramello, Emma and Gallucci, Silvia and Bonato, Marta and Fiocchi, Serena and Parazzini, Marta and Martens, Luc and Joseph, Wout and Ravazzani, Paolo},
  journal={Bioelectromagnetics},
  volume={42},
  number={7},
  pages={550--561},
  year={2021},
  publisher={Wiley Online Library}
}

@inproceedings{falkenberg2018machine,
  title={Machine learning based uplink transmission power prediction for {LTE} and upcoming {5G} networks using passive downlink indicators},
  author={Falkenberg, Robert and Sliwa, Benjamin and Piatkowski, Nico and Wietfeld, Christian},
  booktitle={2018 IEEE 88th Vehicular Technology Conference (VTC-Fall)},
  pages={1--7},
  year={2018},
  organization={IEEE}
}

@article{mazloum2021artificial,
  title={Artificial Neural Network-Based Uplink Power Prediction From Multi-Floor Indoor Measurement Campaigns in {4G} Networks},
  author={Mazloum, Taghrid and Wang, Shanshan and Hamdi, Maryem and Ashenafi Mulugeta, Biruk and Wiart, Joe},
  journal={Frontiers in Public Health},
  volume={9},
  pages={777798},
  year={2021},
  publisher={Frontiers Media SA}
}

@article{egea2019vehicular,
  title={Vehicular networks simulation with realistic physics},
  author={Egea-Lopez, Esteban and Losilla, Fernando and Pascual-Garcia, Juan and Molina-Garcia-Pardo, Jose Maria},
  journal={IEEE Access},
  volume={7},
  pages={44021--44036},
  year={2019},
  publisher={IEEE}
}

@inproceedings{amiot2013pylayers,
  title={{Pylayers}: An open source dynamic simulator for indoor propagation and localization},
  author={Amiot, Nicolas and Laaraiedh, Mohamed and Uguen, Bernard},
  booktitle={2013 {IEEE} International Conference on Communications Workshops (ICC)},
  pages={84--88},
  year={2013},
  organization={IEEE}
}

@inproceedings{mallik2022eme,
  title={{EME-Net}: A {U}-net-based Indoor EMF Exposure Map Reconstruction Method},
  author={Mallik, Mohammed and Kharbech, Sofiane and Mazloum, Taghrid and Wang, Shanshan and Wiart, Joe and Gaillot, Davy P and Clavier, Laurent},
  booktitle={2022 16th European Conference on Antennas and Propagation (EuCAP)},
  pages={1--5},
  year={2022},
  organization={IEEE}
}

@INPROCEEDINGS{11148193,
  author={He, Jingyi and Zheng, Yi},
  booktitle={2025 IEEE/CIC International Conference on Communications in China (ICCC Workshops)}, 
  title={BSCSM-GAN: End-to-End Generative Prediction of Base Station Coverage Strength Maps for 6G-Aware Urban Deployment}, 
  year={2025},
  volume={},
  number={},
  pages={1-5},
  doi={10.1109/ICCCWorkshops67136.2025.11148193}}

\end{document}